\documentclass[pdflatex, sn-nature,Numbered]{sn-jnl}

\usepackage{graphicx}%
\usepackage{multirow}%
\usepackage{amsmath,amssymb,amsfonts}%
\usepackage{amsthm}%
\usepackage{mathrsfs}%
\usepackage[title]{appendix}%
\usepackage{xcolor}%
\usepackage{textcomp}%
\usepackage{manyfoot}%
\usepackage{booktabs}%
\usepackage{algorithm}%
\usepackage{algorithmicx}%
\usepackage{algpseudocode}%
\usepackage{listings}%
\usepackage{soul}

\usepackage{titlesec}
\titlespacing*{\section}
  {0pt}{.5\baselineskip}{.5\baselineskip}
\titlespacing*{\subsection}
  {0pt}{.5\baselineskip}{.5\baselineskip}

\usepackage{tabularx}

\graphicspath{{figures/}}

\begin{document}

\title[Article Title]{The Simons Observatory: Design, Optimization, and Performance of Low Frequency Detectors}


\author*[1]{\fnm{Aashrita} \sur{Mangu}}\email{amangu@berkeley.edu}

\author[1]{\fnm{Benjamin} \sur{Westbrook}}
\author[1]{\sur{Shawn Beckman}}
\author[1]{\sur{Lance Corbett}}
\author[2]{\sur{Kevin T. Crowley}}
\author[3]{\sur{Daniel Dutcher}}
\author[4]{\sur{Bradley R. Johnson}}
\author[1,5]{\sur{Adrian T. Lee}}
\author[1]{\sur{Varun Kabra}}
\author[1]{\sur{Bhoomija Prasad}}
\author[3]{\sur{Suzanne T. Staggs}}
\author[5]{\sur{Aritoki Suzuki}}
\author[3]{\sur{Yuhan Wang}}
\author[3]{\sur{Kaiwen Zheng}}

\affil*[1]{\orgname{University of California, Berkeley}, \orgaddress{\city{Berkeley}, \state{CA}, \country{USA}}}

\affil[2]{\orgname{University of California, San Diego}, \orgaddress{\city{San Diego}, \state{CA}, \country{USA}}}

\affil[3]{\orgname{Princeton University}, \orgaddress{\city{Princeton}, \state{NJ}, \country{USA}}}

\affil[4]{\orgname{University of Virginia}, \orgaddress{\city{Charlottesville}, \state{VA}, \country{USA}}}

\affil[5]{\orgname{Lawrence Berkeley National Laboratory}, \orgaddress{\city{Berkeley}, \state{CA}, \country{USA}}}


\abstract{The Simons Observatory (SO) is a cosmic microwave background (CMB) experiment located in the Atacama Desert in Chile that will make precise temperature and polarization measurements over six spectral bands ranging from 27 to 285 GHz. Three small aperture telescopes (SATs) and one large aperture telescope (LAT) will house $\sim$60,000 detectors and cover angular scales between one arcminute and tens of degrees. We present the performance of the dichroic, low-frequency (LF) lenslet-coupled sinuous antenna transition-edge sensor (TES) bolometer arrays with bands centered at 27 and 39 GHz. The LF focal plane will primarily characterize Galactic synchrotron emission as a critical part of foreground subtraction from CMB data. We will discuss the design, optimization, and current testing status of these pixels.}




\maketitle

\section{Introduction}\label{sec1}

Cosmic microwave background (CMB) experiments have paved the way towards a greater understanding of early universe physics and building a standard model of cosmology. The Simons Observatory (SO) will make multi-frequency CMB maps from the Atacama Desert in Chile. SO is nominally aiming for science observations starting in 2024 with  $\sim$60,000 detectors to perform degree-scale surveys with three small aperture telescopes (SATs) and arcminute-resolution surveys with the large aperture telescope (LAT). These telescopes are specifically designed to reach key science targets including constraints of the tensor-to-scalar ratio \textit{r}, the effective number of relativistic species $N_{eff}$, the sum of neutrino masses $\Sigma m_{\nu}$, deviations from the cosmological constant, galaxy evolution feedback efficiency, and the epoch of reionization \cite{SOforecasting}. Galactic and extra-Galactic foreground-subtracted CMB data are critical to achieving these science goals \cite{smoot1999, wmap_2009, wmap_2009_2, foregrounds2016, Sponseller_2022, SOforecasting, cmbs4science, Kamionkowski_2016, planck2018_iv}. To achieve sensitivities that enable this science, SO will be observing across six spectral bands from 27 to 285 GHz. Focal planes across the telescopes will be dichroic, observing at the Low-Frequency (LF) bands centered at 27 and 39 GHz, the Mid-Frequency (MF) primary CMB bands centered at 93 and 145 GHz, and the Ultra-High-Frequency (UHF) bands centered at 225 and 285 GHz. The UHF bands will characterize thermal dust emission and the LF bands will characterize synchrotron emission, which are the dominant emission sources in Galactic polarization signals \cite{planck2018_1, Planck2018_xi, planck2018_iv}. Here, we focus on the LF focal plane development studying primarily Galactic synchrotron emission. 

\section{Design and Optimization}\label{sec2}

\subsection{Overview: Optical Stack and Pixel Design}

All SO telescopes have hexagonally tiled focal planes. Each hexagonal tile is called a Universal Focal Plane Module (UFM), and houses optical and readout components. UFMs are interchangeable (i.e. ``universal") between any of the SATs and LAT. For each LF-UFM, a Detector and Optical Stack (DOS) mates with a Universal Mutiplexing Module (UMM). Each LF-UFM consists of 37 pixels, with 4 optical and 2 dark transition-edge sensor (TES) detectors per pixel. Each of the 4 optical TESs covers one of the two bands and one of two orthogonal polarizations. Only one of each pair of dark detectors is bonded to be read out. The UMM consists of a routing wafer and multiplexer chips to enable the microwave multiplexing scheme that reads out $O(1000)$ detectors per readout channel \cite{mccarrick2021, Healy_2022}. The LF-DOS design includes an interlocking flange to mate with other UFMs on the focal plane, the lenslet array on the detector wafer, the copper backshort, and an array support structure (see Figure \ref{fig1}).

The SO LF pixel arrays consist of lenslet-coupled sinuous antennas. The borosilicate glass anti-reflection (AR) coated lenslets help focus the incoming light onto the antenna and maximize its forward gain. The ultra-wide-band dual-polarized sinuous antenna covers 27 and 39 GHz bands with total $\sim$75\% fractional bandwidth, with diplexers splitting bands with 32\% and 44\% fractional bandwidth respectively.

The sinuous antenna couples to superconducting niobium (Nb) microstrip lines, which carry the signal to diplexing lumped element bandpass filters. After passing cross-unders, the signal is terminated at a TES bolometer on a titanium (Ti) load resistor. (See Figure \mbox{\ref{fig2}} for details.) Palladium (Pd) thermal ballasts control the time constant of the bolometer. The TES target requirements are a transition temperature ($T_C$) of 160 mK, normal resistance ($R_n$) of 8 m$\Omega$, and saturation powers ($P_{sat}$) of 0.83 pW [27 GHz, optical], 3.54 pW [39 GHz, optical], and 2.07 pW [dark] to be consistent with SO sensitivities \cite{SOforecasting}. The aluminum manganese (AlMn) TESs are tuned to their superconducting transition and operated by the warm microwave multiplexing readout electronics developed by SLAC \cite{Henderson_2018, Kernasovskiy_2018, Yu_2023} with carefully designed readout chains \cite{mayuri_2020}). 

\subsection{Optimization}

\begin{figure}[t]%
\centering
\includegraphics[width=\textwidth]{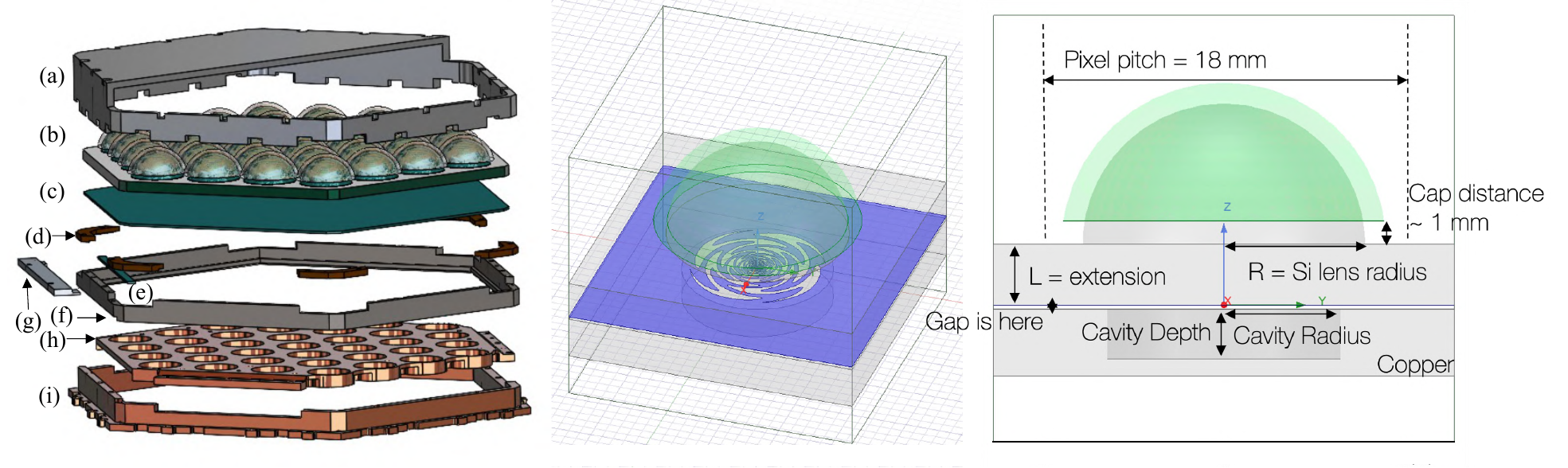}
\vskip-0.5ex
\caption{\textit{Left:} DOS CAD exploded view. Components are: \textbf{(a)} vacuum plate**; \textbf{(b)} LF pixel array; \textbf{(c)} detector wafer; \textbf{(d)} invar clips; \textbf{(e)} bond-pad piece; \textbf{(f)} LF array support structure; \textbf{(g)} step-up aligner piece**; \textbf{(h)} copper backshort piece; and \textbf{(i)} interlocking flange. Components labelled ** were used for shipping/assembly and then removed. \textit{Center:} Single pixel simulation trimetric view. Blue is ground plane, green is AR coating. \textit{Right:} Simulation front view with key parameters labels }\label{fig1}
\vskip-2ex
\end{figure}

\begin{table}[b]
\vskip-2ex
\caption{Optimized Pixel Parameters}\label{tab1}%
\vskip-1ex
\begin{tabularx}{\textwidth}{ >{\hsize=.8\hsize}X | X }
    \hline
    \textbf{Parameter} & \textbf{Design Value/ Consideration} \\
    \hline
    Cavity Radius  & 6.5 mm \\
    \hline
    Cavity Depth  & 3 mm \\
    \hline
    Gap between extension and backshort  & 250 um
    [$\leq$2\% variation between 50-750um] \\
    \hline
    L/R ratio  & 0.43 \\
    \hline
    Swiss Cheese backshort  & performs comparably to standard flat backshort \\
    \hline
\end{tabularx}
\vskip-0.5ex
\end{table}

The pixel design was optimized using ANSYS HFSS simulations. As shown in Figure \ref{fig1}, the geometry consists of a hemispherical lenslet with AR coating, the device and extension wafer, and the copper backshort. A new AR coating material using borosilicate glass was used and compared to traditional epoxy coatings \cite{suzuki2012, suzuki2016, Westbrook_2018}, which will be discussed in an upcoming paper. A primary goal of these studies was to optimize the backshort design. Limited space from UFM design required a compact backshort to accommodate the UMM readout hardware. As a solution, the `swiss cheese' backshort was designed to replace the traditional flat backshort sitting $\lambda_{center}/4$ from the antenna plane, where $\lambda_{center}$ corresponds to the central frequency of the combined 27 and 39 GHz bands. It consists of a metal slab with a small cavity behind antenna to recover power transmitted in the backlobe. Given the silicon half-space and AR-coated lenslet sitting above the antenna itself, less than 5\% of the signal sits in the back. We aimed to perform comparably to or better than the flat backshort, and the simulated forward integrated gain suggests this new backshort performs better by $\sim$1\%.

\begin{figure}[b]%
\vskip-2ex
\centering
\includegraphics[width=\textwidth]{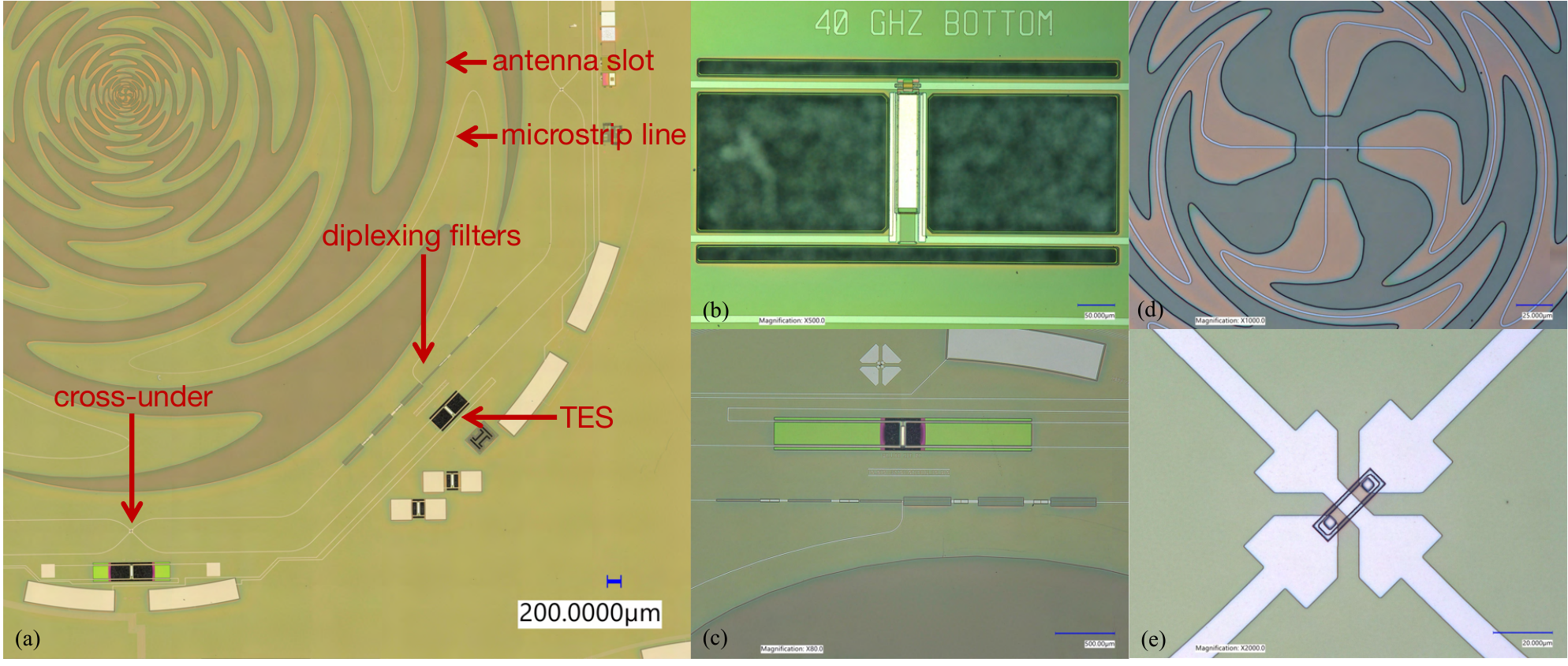}
\vskip-1ex
\caption{SO LF wafer images taken with Keyence 7000. \textbf{(a)} Pixel view showing sinuous antenna, superconducting microstrip line, lumped element band pass filters, cross-unders, and TES. \textbf{(b)} 39 GHz  TES bolometer (labelled as approximate band center, 40 GHz). Visible are microstrip lines differentially feeding a Ti load resistor. Pd thermal ballasts break the structure from the AlMn TES on the top end, which is also connected to two microstrip lines. \textbf{(c)} Zoomed in view of end of sinuous slot to signal diplexing into both 27 and 39 GHz band pass filters. 27 GHz bolometer is also visible. \textbf{(d)} Sinuous antenna center fed by two microstrip lines. This is a new, updated center design to maintain self-complementarity compared to previous designs \cite{suzuki2012, suzuki2016, Westbrook_2018}. \textbf{(e)} Cross-unders allow microstrip lines to cross over each other to reach TESs. Widening of the microstrip line near the via minimizes loss}\label{fig2}
\vskip-0.5ex
\end{figure}

Each pixel in the array is backed by a single cavity. The cavity radius and depth were optimized by using iterative sweeps in simulation and choosing values with maximum band-averaged forward integrated gain (in the skyward direction) first between $\pm30^{\circ}$ for coarse cuts, and then more finely between $\pm13^{\circ}$ for the LAT and $\pm17^{\circ}$ for the SAT Lyot stop \cite{ali2020, galitzki2018, harrington2020, Zhu_2021}.

The gap between the extension wafer and the copper backshort piece was investigated to see how deviations from a 250um gap would also affect pixel performance. We saw less than 2\% variation in forward integrated gain for gap sizes between 50 and 750um. Cross-talk effects are considered negligible with changes in gap dimension, because (1) the propagation space here is much smaller than if a flat backshort is used and (2) very little power is in the backlobe, as stated earlier.

The ratio of the silicon extension (L) to the radius of the silicon lenslet (R) is defined as the L/R ratio. This parameter tunes how effectively the antenna is the focal point of the effective ellipsoid made by the lenslet plus extension geometry. We tuned this parameter to optimize forward integrated gain and beam shape.

The conclusion of these optimizations was that the swiss cheese backshort performs better or comparably to a standard flat backshort (see Figure \ref{fig5}(a)). A summary of final design values used are listed in Table \ref{tab1}. 

\section{Fabrication and Ongoing Testing}\label{sec3}

\begin{figure}[t]%
\centering
\includegraphics[width=\textwidth]{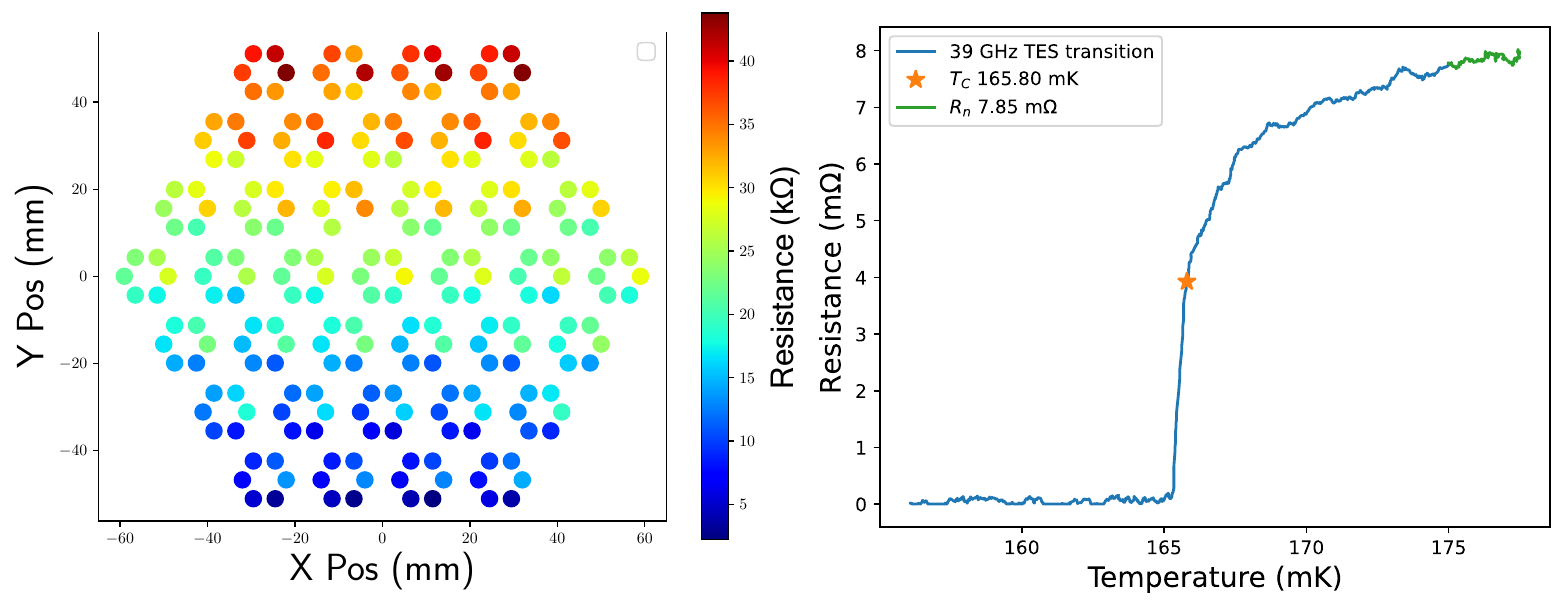}
\vskip-.5ex
\caption{\textit{Left:} Warm resistance map, dominated by the gradient in niobium wire lengths used to bias TESs (which turn superconducting at operational temperatures). These maps identify any opens or shorts in the circuit. We achieved $\sim$96\% warm electrical yield on this wafer. \textit{Right:} Transition within specifications with Rn $\approx$ 8 m$\Omega$ and Tc $\approx$ 165 mK}\label{fig3}
\vskip-2ex
\end{figure}

\begin{figure}[b]%
\vskip-2ex
\centering
\includegraphics[width=\textwidth]{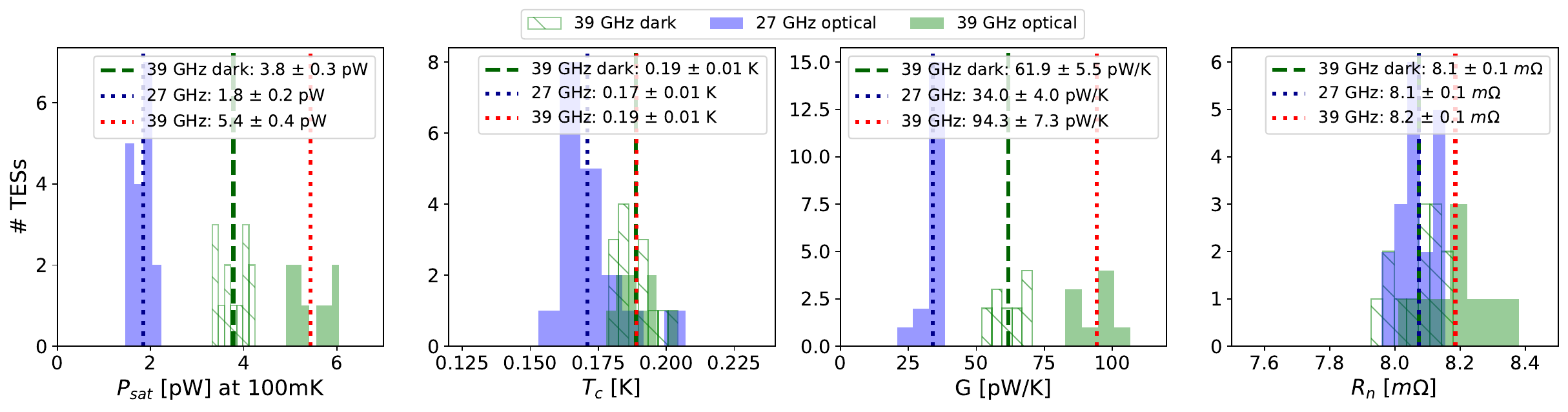}
\vskip-1ex
\caption{Thermal parameter distributions for a tested half-wafer with a cold load. Numbers reported are mean with 1$\sigma$ errors. Note that dark bolometers have lower $P_{sat}$ by design. Similarly, ``optical" refers to all antenna-coupled bolometers. \textit{Left:} Saturation powers ($P_{sat}$). \textit{Center Left:} Transition temperatures ($T_c$) \textit{Center Right:} Thermal conductances ($G$). \textit{Right:} Normal resistance ($R_n$)} 
\vskip-0.5ex
\label{fig4}
\end{figure}  

\begin{figure}[t]%
\centering
\includegraphics[width=\textwidth]{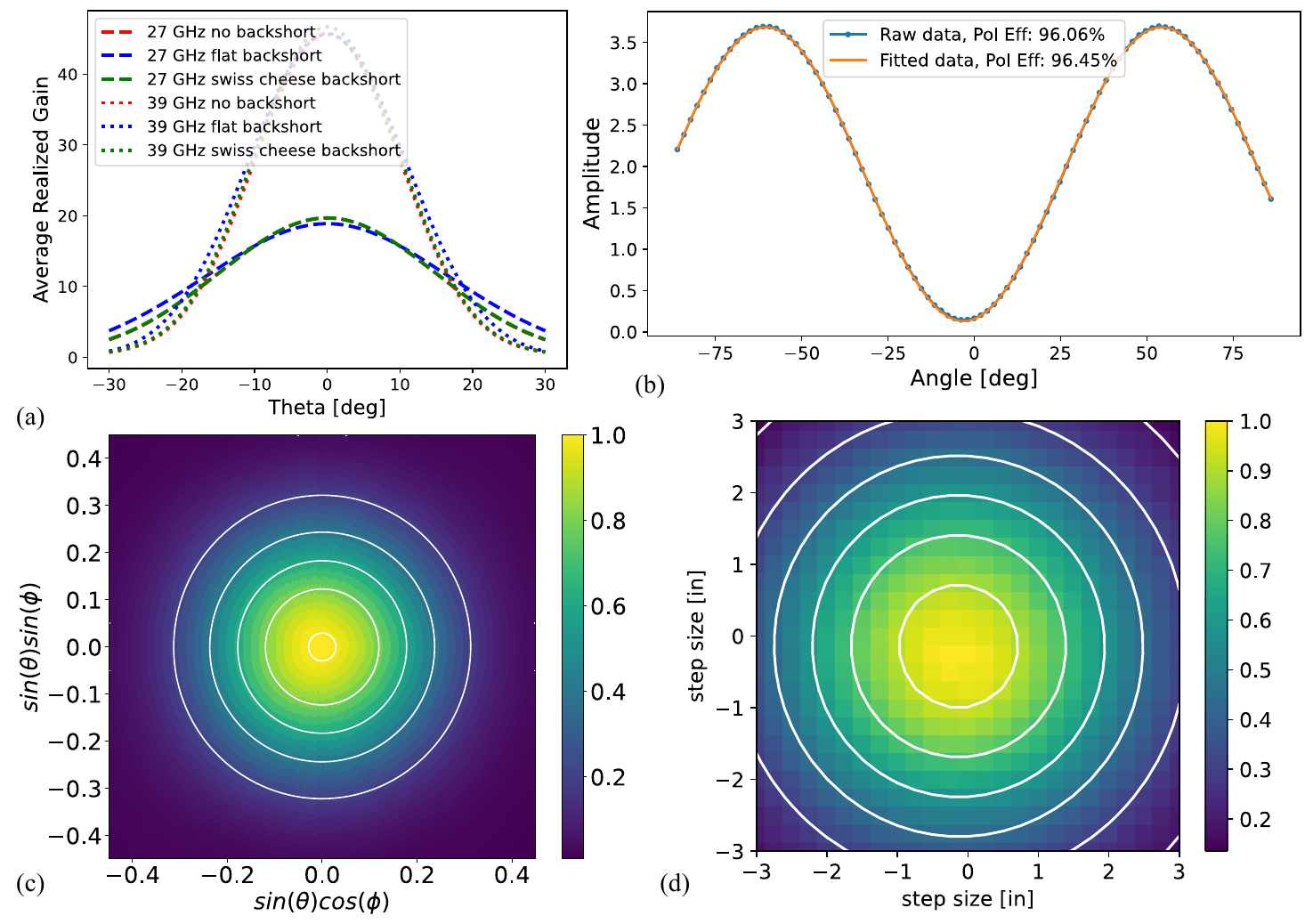}
\vskip-0.5ex
\caption{\textbf{(a)} `Swiss cheese' backshort performs better than a flat backshort in beam width and amplitude for both bands in simulation. Note that the `27 GHz no backshort' line is hidden behind the `27 GHz swiss cheese backshort' line. \textbf{(b)} Polarization efficiency was measured using a rotating polarizing wire grid, and is $>$96\%. \textbf{(c)} and \textbf{(d)} Left is simulated 39 GHz band beam using ANSYS HFSS. Right is 39 GHz band measured beam using a beam mapper. Contours are 2D gaussian fits, with $\sim1\%$ ellipticity for both simulated and measured beams}\label{fig5}
\vskip-1.5ex
\end{figure}

Fabrication of the LF array is done at the Marvell Nanofabrication Laboratory at UC Berkeley (see Figure \ref{fig2}). The fabrication flow is adapted from the fabrication flow used by UC Berkeley for the Simons Array focal planes \cite{suzuki2012, suzuki2016, Westbrook_2018, westbrook2022}. The primary challenge to fabricating these devices is to realize the bolometric and radio frequency (RF) features simultaneously. Fabrication begins with 6 inch silicon wafers with a $\sim$600~nm film of low-pressure chemical vapor deposition low-stress silicon nitride grown on both sides of the wafer in a furnace. The base of the RF circuit is created when a patterned layer of $\sim$340~nm thick Nb is coated with $\sim$1.0~$\mu$m of silicon nitride. The nitride serves as the dielectric layer for the RF circuitry. Next, the Ti load resistor and AlMn TES are deposited, patterned and etched. The Ti resistor is $\sim$80~nm thick and the AlMn TES is $\sim$400~nm thick. An encapsulating layer of passivating silicon nitride is patterned over the load resistor and TES to protect these layers from the subsequent Nb etch and serves as a long-term passivation layer for the devices. 

The RF circuit is then completed with the deposition and etching of a second Nb layer that forms both the RF microstrip line and TES wiring layers. This wiring layer also provides access to the readout with bond pads at the wafer edge. We then evaporate $\sim$500~nm of Pd on to the bolometer islands using a lift-off process. At this point, the wafer is ready for release and packaging. A layer of photoresist is patterned, allowing access to the silicon nitride layer that will eventually support the suspended bolometer arms. This nitride is etched before the wafer is diced into its final hexagonal shape with chamfers at the corners. The bolometers are released with xenon difluoride (XeF$_{2}$) vapor providing thermal isolation of the TES from the rest of the substrate. With a final photoresist ash, the wafer is ready for inspection and characterization. 

Current testing result highlights are shown in Figures \ref{fig3}, \ref{fig4}, and \ref{fig5}. Thermal parameters are within internal requirements for deployment, and wafers have good warm electrical yields. Ongoing optical testing show good polarization efficiency, and beam measurements have low ellipticity and are consistent with simulated beam profiles. Full SO LF testing results will be detailed in an upcoming paper.

\section{Current Status and Conclusions}\label{sec4}

SO LF wafer fabrication and testing is ongoing as described in Section \ref{sec3}, with an optimized swiss cheese backshort design, good TES thermal parameters, warm electrical yield, polarization efficiency, and round beams. The LF detectors will primarily characterize synchrotron emission, an important foreground to understand and subtract from CMB data to achieve key science targets. The LAT will contain 444 LF detectors on the sky, with a future SAT to put an additional 1,036 LF detectors on the sky. SO is currently deploying the LAT and three SATs, and full science observations will begin in 2024.

\backmatter
\bmhead{Acknowledgments} This work was supported in part by the Simons Foundation (Award \#457687, B.K.).

\bibliography{sn-bibliography}

\end{document}